\documentclass[twocolumn,showpacs,preprintnumbers]{revtex4-1}
\usepackage{graphicx}
\usepackage{dcolumn}
\usepackage[T1]{fontenc}
\usepackage[english]{babel}
\usepackage{amsmath}
\usepackage{amssymb}
\usepackage{tabularx}
\usepackage{multirow}
\usepackage[all]{xy}
\usepackage{appendix}
\usepackage{xspace}
\usepackage{color}

\newcommand{\ket}[1]{\ensuremath{|#1\rangle}\xspace}
\newcommand{\bra}[1]{\ensuremath{\langle #1|}\xspace}

\begin{document}
\title{High-field quantum calculation reveals time-dependent negative Kerr contribution}

\author{P. B\'ejot$^{1}$} \email{pierre.bejot@u-bourgogne.fr}
\author{E. Cormier$^{2}$}
\author{E. Hertz $^{1}$}
\author{B. Lavorel$^{1}$}
\author{J. Kasparian$^{3}$}
\author{J.-P. Wolf$^{3}$}
\author{O. Faucher$^{1}$}

\affiliation{$^{1}$ Laboratoire Interdisciplinaire CARNOT de Bourgogne, UMR 6303 CNRS-Universit\'e de Bourgogne, BP 47870, 21078 Dijon, France}
\affiliation{$^{2}$ Centre Lasers Intenses et Applications, Universit\'e de Bordeaux-CNRS-CEA, UMR 5107, 351 Cours de la Lib\'eration F-33405 Talence,
France} \affiliation{$^{3}$ Universit\'e de Gen\`eve, GAP-Biophotonics, Chemin de Pinchat 22, 1211 Geneva 4, Switzerland}

\begin{abstract}
The exact quantum time-dependent optical response of hydrogen under strong field near infrared excitation is investigated and compared to the perturbative model widely used for describing the effective atomic polarization induced by intense laser fields. By solving the full 3D time-dependent Schr\"{o}dinger equation, we exhibit a supplementary, quasi-instantaneous defocusing contribution missing in the weak-field model of polarization. We show that this effect is far from being negligible in particular when closures of ionization channels occur and stems from the interaction of electrons with their parent ions. It provides an interpretation to higher-order Kerr effect recently observed in various gases.
\end{abstract}

 \maketitle

Atoms and molecules exposed to a nonresonant intense (10-100 TW/cm$^2$) laser field exhibit highly nonlinear dynamics that has motivated a wealth of experimental and theoretical studies and led to the observation of phenomena such as above-threshold ionization (ATI) \cite{Eberly91}, high-harmonics generation (HHG) \cite{Ferray88}, attosecond pulses generation \cite{Antoine96}, or filamentation \cite{ChinHLLTABKKS05,BergeSNKW07,CouaironM07,KasparianW08}. Understanding the first three processes has required to describe the atomic and molecular dynamics under strong-field excitation. In this regime, the electric field is as intense as the intra-atomic field, so that electronic transitions can be enhanced by dynamic resonances leaving the atom in a coherent superposition of bound and continuum states. In the same time, ionization channel closures occur, i.e. the minimum number of photons required to ionize the atom increases by several units due to the Stark shift of the ionization potential \cite{Cormier2}. Because of this complex dynamics, the atomic optical response can no longer be described as a perturbative series of the field. Instead, the atom dynamics and its associated polarization must be evaluated at each time during the interaction by solving the time-dependent Schr\"{o}dinger equation (TDSE) describing the interacting atom.
As far as the filamentation process is concerned, it is generally interpreted through a perturbative  approach as resulting from a dynamic balance between Kerr self-focusing and defocusing by the free electrons originating from ionization during the interaction (Drude model). Retrospectively, it is surprising that this process is still described in the perturbative framework, contrary to HHG or ATI. Indeed, since the typical intensity in filaments (50 TW/cm$^2$ \cite{KaspaSC2000,BeckeAVOBC2001}) is of the same order as that used in HHG or ATI experiments, it seems natural to wonder whether filamentation can still be accurately described within the lowest order perturbation theory, i.e. using the commonly used Kerr and Drude (KD) model. In particular, the saturation and even the inversion  of the Kerr effect (i.e. the sign inversion of the nonlinear refractive index), empirically described as negative higher-order Kerr effect (HOKE), were reported between {19 and 33 TW/cm$^2$} in major air components \cite{LoriotHFL09,LaserPhysLoriot}. A laser-induced transient grating \cite{Milchberg} was also proposed as an alternative interpretation of the measurements, casting doubts on the relevance of this model. Nevertheless, saturation and inversion of the Kerr index, predominantly attributed to continuum-continuum transitions, were predicted in hydrogen \cite{Kano,Nurhuda1,Nurhuda2}, in atomic silver \cite{Volkova}, in H$_2^+$ \cite{Bandrauk}, as well as with TDSE calculations in a one-dimensional Dirac potential \cite{TelekiWK2010}, or by considering Kramers-Kronig relations and multiphoton ionization rates \cite{BreeDS2011}.   

In this Letter, we  perform 3D non-perturbative ab-initio calculations of the interaction between a near infrared strong laser field and an hydrogen atom. The choice of hydrogen is motivated by the possibility of performing rigorous exact calculations without any assumption, in particular about the exact form of the potential. We show that the usual perturbative model of polarizability, describing a purely third-order Kerr medium together with a negative plasma contribution following the Drude model, misses a substantial additional negative contribution associated with ionization channel closures. The latter contribution is due to the over-acceleration of the electrons interacting with their parent ions during the pulse and can be identified as the empirically-introduced higher order Kerr effect \cite{LoriotHFL09,LaserPhysLoriot}. This effect may affect any strong-field experiment in long gaseous media. The Kerr effect is further influenced by field-induced resonances that eventually leave several excited states significantly populated at the end of the pulse. Moreover, we show that the Kerr inversion mechanism cannot be specifically attributed to neither bound-bound nor continuum-continuum transitions that are gauge dependent. This stems from the fact that the atomic states are not eigenstates of the dressed atomic system, and consequently are not quantum observables of the system during the interaction. This result sheds a new light on the higher order Kerr effect controversy which had crystalized the debate around this question \cite{Kano,Nurhuda1,Nurhuda2,Volkova, Bandrauk}. 

Within the dipole approximation, the TDSE describing the electron wavefunction $\ket{\psi}$  evolution in the presence of an electric field $\textbf{E}(t)$ linearly polarized along the axis $z$ reads:

\begin{equation}
i\frac{d\ket{\psi}}{dt}=(H_0+H_{\textrm{int}})\ket{\psi},
\end{equation}
where $H_0=\mathbf{\nabla}^2/2-1/r$ is the hydrogen atom Hamiltonian, $H_{\textrm{int}}=-\textbf{E}(t)\cdot\textbf{r}$ (resp. $H_{\textrm{int}}=\textbf{A}(t)\cdot\boldsymbol{\pi}$, where $\textbf{A}(t)$ is the vector potential such that $\textbf{E}(t)=-\partial \textbf{A}/\partial t$ and $\boldsymbol{\pi}=-i\mathbf{\nabla}$) is the interaction term expressed in the length gauge (resp. velocity gauge).
The time-dependent wavefunction $\ket{\psi}$ is expanded on a finite basis of B-splines  \cite{Bachau2001} allowing memory efficient fast numerical calculations with a very large basis set:

\begin{equation}
  \psi(\textbf{r},t) = \sum_{l=0}^{l_\textrm{max}}\sum_{i=1}^{n_\textrm{max}} c_{i}^{l}(t)
  \frac{B_{i}^k(r)}{r}Y_{l}^{0}(\theta,\phi),
\label{eq:TDSE_expansion}
\end{equation}
where $B_i^k$ and $Y_l^m$ are B-spline functions and spherical harmonics, respectively. The basis parameters ($l_\textrm{max}$, $n_\textrm{max}$, $k$ and the spatial box size) and the propagation parameters are chosen to ensure convergence \cite{CormierGauge}. Unless otherwise specified, calculations are performed in the velocity gauge where computation is eased \cite{CormierGauge}. The atom is initially in the ground state (1s) and the electric field $E$ is expressed as $\textrm{E}(t)=E_0\cos(t/\sigma_\textrm{t})^2\cos(\omega_0t)$, where $\omega_0$ is the central pulsation of the laser, $\sigma_\textrm{t}=2N/\omega_0$, and $N$ the total number of optical cycles in the temporal window. The simulations are performed for a laser wavelength of 800 nm and pulse durations (FWHM of intensity) of 23, 29, 47, 58, and 93 fs (corresponding to  $N$=24, 30, 48, 60, and 96 cycles respectively).

 \begin{figure}[tb!]
\begin{center}
      \includegraphics[keepaspectratio,width=8.5cm]{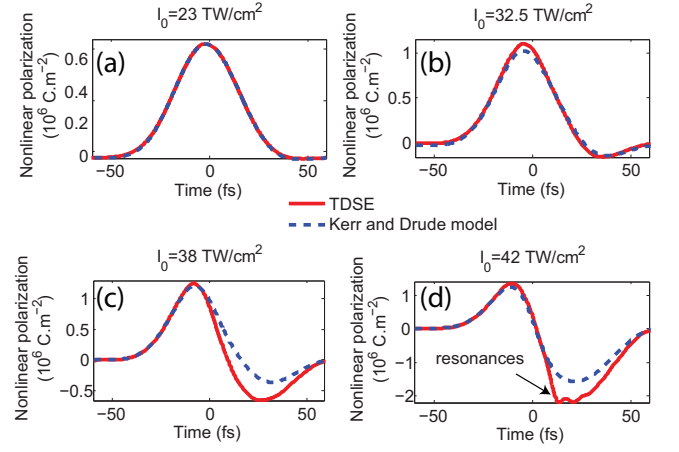}
\end{center}
   \caption{(Color online) Nonlinear polarization envelope as a function of time for a 30 (a), 32.5 (b), 38 (c), and 42 (d) TW/cm$^2$ 48 cycles pulse.}
\label{Fig:1}
\end{figure}

The microscopic polarization $p(t)$ and the medium polarizability $\alpha$ at the driving frequency $\omega_0$ are defined in atomic units as:
\begin{eqnarray}
p(t)&=&\bra{\psi(t)}r\cos\theta\ket{\psi(t)}\\ 
\alpha(\omega_0)&=&\frac{p(\omega_0)}{E(\omega_0)}, 
\end{eqnarray}
where $p(\omega)$ (resp. $E(\omega)$) is the Fourier transform of $p(t)$ (resp. $E(t)$) and $\theta$ is the angle between the dipole moment and its projection along the electric field. No collective effect is expected at the considered time scale \cite{MoloneyOpex}, even if a comprehensive treatment would need to consider the interaction with the potential of the neighboring atoms, which is beyond the scope of the present work. Under the isolated atom assumption, the macroscopic polarization $P$ is then calculated as $P(t)=\mathcal{N}p(t)$, with $\mathcal{N}$ the number density. 

\begin{figure}[tb!]
\begin{center}
      \includegraphics[keepaspectratio,width=8.5cm]{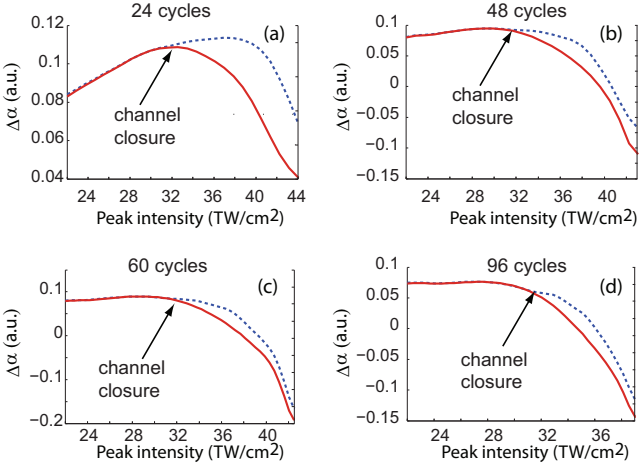}
\end{center}
   \caption{(Color online) Nonlinear polarizability as a function of peak intensity for 24 (a), 48 (b), 60 (c) and 96 (d) cycles pulse calculated with TDSE (red solid line) and KD model (blue dashed line), respectively.}
\label{Fig:2}
\end{figure}

Its nonlinear part $P_{\textrm{NL}}$ and the nonlinear polarizability $\Delta\alpha$ are calculated as:
\begin{eqnarray}
	P_{\textrm{NL}}(t)&=&P(t)-\mathcal{N}\lim_{I_0 \ \mapsto 0 } \limits \alpha(I_0)E(t),\\
	 \Delta\alpha(I_0)&=&\alpha(I_0)-\lim_{I_0 \ \mapsto 0 } \limits \alpha(I_0), 
\end{eqnarray}
where $I_0$ is the peak intensity of the pulse.
As shown in Fig. \ref{Fig:1}, the nonlinear polarization envelope (the carrier and the harmonics have been filtered out) follows the intensity profile in the low-field regime ($I_0$<30 TW/cm$^2$), as expected for a purely cubic Kerr effect. As the intensity increases, a transient, time-retarded negative contribution develops on the second half of the pulse and leads to a sign inversion of the nonlinear polarization in the falling edge of the pulse. In the frequency domain, the nonlinear polarizatiblity of the atom at the fundamental frequency $\omega_0$ increases linearly with the pulse peak intensity, saturates, and eventually becomes negative (Fig. \ref{Fig:2}). The threshold intensity for inversion is in line with previous works \cite{Nurhuda1,Nurhuda2,Kano,Bandrauk,TelekiWK2010} and similar for all investigated pulse durations.

\begin{figure}[tb!]
\begin{center}
      \includegraphics[keepaspectratio,width=8.5cm]{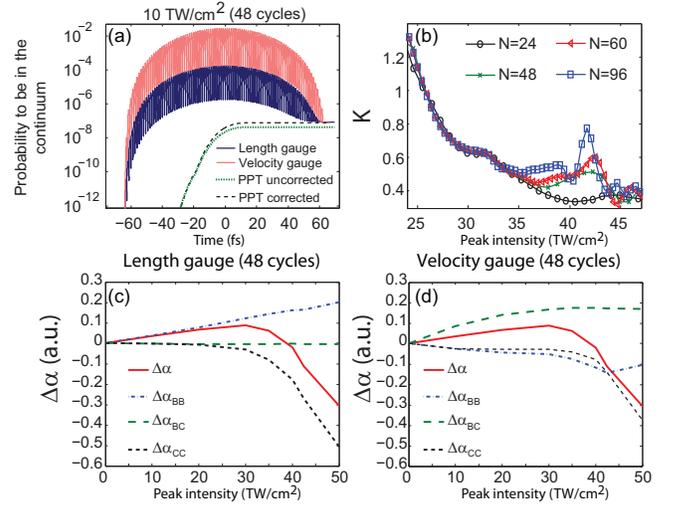}
\end{center}
   \caption{(Color online) (a) Probability of the electron to be in the continuum during a 48 cycles 10 TW/cm$^2$ pulse calculated in the length and velocity gauges, respectively, compared with the output of the PPT formula. (b) Correction factor $K$ of the PPT ionization probability. Nonlinear polarizability as a function of intensity calculated in (c) the length and (d) velocity gauge, respectively. $\Delta\alpha_\textrm{BB}$, $\Delta\alpha_\textrm{BC}$, and $\Delta\alpha_\textrm{CC}$ refer to the partial nonlinear polarizabilities related to bound-bound, bound-continuum, and continuum-continuum transitions, respectively.}
\label{Fig:3}
\end{figure}

From this point, one can wonder if the usual way to describe the atomic polarizability at the fundamental frequency through a perturbative approach still remains valid at high intensity or if supplementary effects must be considered. To answer we shall first recall that the KD model of polarization is based on the following assumptions: i) atoms are responsible for a pure cubic Kerr effect,
ii) free electrons are produced through ionization during the interaction with the field while depletion of the ground state population remains negligible,
iii) the ionized electrons (commonly called plasma), insensitive to the atomic potential, accelerate in the field as free particles and are responsible for balancing the Kerr effect.
Following the above assumptions, the effective nonlinear macroscopic polarization of the system reads:
\begin{equation}
	P_{\textrm{NL,KD}}^{\textrm{eff}}(t)=\epsilon_0\left(n_2I(t)-\frac{\rho(t)}{\rho_c}\right)E(t),	\label{modele_classique}
\end{equation}
where $\rho$ is the free electrons density, 
$\rho_\textrm{c}=m_\textrm{e}\omega_0^2\epsilon_0/q^2$ is the critical plasma density, $n_2$ is the intensity independent nonlinear index of the medium (which can be extracted from TDSE calculations at weak intensity), $\epsilon_0$ is the vacuum permittivity, and $m_\textrm{e}$ and $q$ are the electron mass and charge, respectively. Since Eq. \ref{modele_classique} depends on the free electrons density, the perturbative scenario de facto assumes that the latter is a quantum observable during the interaction, and, as such, can be measured and evaluated. In that framework, models like multiphoton ionization, PPT, or ADK can be used to evaluate the ionization probability during the interaction \cite{Keldysh,PPT,ADK}. 
It should be emphasized that this approach is inappropriate when dealing with quantum mechanics because the wavefunction describing the system depends on the representation of the interaction (i.e. the gauge), whereas the physical predictions (for instance, the electron position and mechanical momentum) do not. 
As shown in Fig. \ref{Fig:3}(a), the probability for the electron to be in the continuum of the unperturbed atom (corresponding to the ionized electrons density) varies by several decades during the interaction depending on the considered gauge and therefore is not a quantum observable. The ionized electron density cannot therefore be measured during the interaction, making the time-dependent KD scenario questionnable from a quantum mechanical point of view. To show the divergence between the full TDSE calculations and the KD model, we used the Perelomov-Popov-Terent'ev (PPT) model \cite{PPT} for evaluating $\rho$. Since PPT overestimates the postpulse electron densities calculated with TDSE, which gives exact results in the case of atomic hydrogen, we corrected it so that $\rho(t=+\infty)=\rho_{\textrm{TDSE}}(t=+\infty)=K \rho_{\textrm{PPT}}(t=+\infty)$, where $K$ is an intensity dependent correction factor depicted in Fig. \ref{Fig:3}(b). Note that, the amount of ionized electrons becomes an appropriate quantum observable after extinction of the external electric field  [see Fig. \ref{Fig:3}(a)].

The KD model qualitatively reproduces the nonlinear polarization dynamics calculated with TDSE (Figs. \ref{Fig:1} and \ref{Fig:2}). However, it significantly under-estimates the negative contribution leading to the sign inversion of the nonlinear polarizability. As shown in Fig. \ref{Fig:2}, the discrepancy between the two models develops as soon as the intensity exceeds the 10-photons ionization channel closure ($I_{0} \simeq$ 32 TW/cm$^2$, see Fig. \ref{Fig:2}) where 11 photons are necessary to ionize the atom as the ionization threshold shifts upward due to the AC stark effect, and further increases above the 11-photons ionization channel closure at $I_{0}=$57 TW/cm$^2$. Ionization channel closure turns out to be a key ingredient of the negative nonlinear transient contribution to the polarization. Its impact appears similar to that induced by the ionization suppression mechanism \cite{Ivanov} but takes place at lower intensities. 

\begin{figure}[tb!]
\begin{center}
      \includegraphics[keepaspectratio,width=8.5cm]{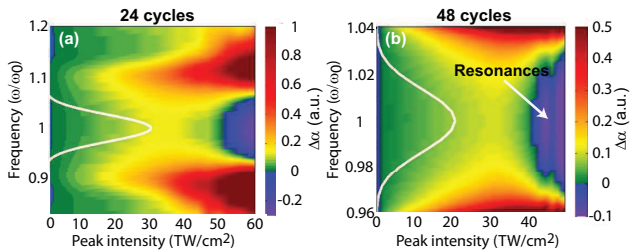}
\end{center}
   \caption{(Color online) Spectral dependence of the nonlinear polarizability for (a) 24 and (b) 48 cycles.}
\label{Fig:Spectral}
\end{figure}

During the interaction, the KD model of polarization deviates from the quantum calculations with the fourth power of the incident peak intensity, consistent with its empirical identification with  a negative $n_8 I_0^4$ contribution to the refractive index \cite{LoriotHFL09} at large intensity. The negative effect induced by ionization channel closures is therefore similar (except for the delay of few femtoseconds, compatible with the experimental temporal resolution), to the higher order Kerr effect introduced empirically to describe the experimental observations of \textit{intra-pulse} Kerr saturation and inversion \cite {LoriotHFL09,LaserPhysLoriot}. The major air constituents, as well as most usual gases, exhibit ionization channel closures in the same intensity range as hydrogen (21.5 and 48 TW/cm$^2$ in Ar, 32 and 57 TW/cm$^2$ in O$_2$, and 25 and 51 TW/cm$^2$ in N$_2$). As a consequence, in all of these gases, the effect of ionization channel closures has to be taken into account in the sign reversal of the nonlinear refractive index. It should be noted that the nonlinear polarizability saturates and reverses its sign only at frequencies within the bandwidth of the incident pulse spectrum (see Fig. \ref{Fig:Spectral}) as recently observed in \cite{Levis2Couleurs}. This unexpected finding is consistent with the apparent contradiction between studies performed at a single wavelength, which exhibit refractive index saturation and inversion \cite{LoriotHFL09,LaserPhysLoriot,Levis2Couleurs}, and two-color investigations, that do not \cite{Levis2Couleurs,MilchbergPRL,Molo2Couleurs}.

Beyond the ionization channel closure, excited states dynamically shift with intensity. The resulting dynamical resonances lead to a net  population transfer to some excited states after the pulse turn-off. For instance, 2d, 3d, and 1-4g states are signifiquantly  populated after the interaction with a pulse of  $\simeq$40 TW/cm$^2$ peak intensity, impacting the atomic polarization [Fig. \ref{Fig:1}(d)] and the associated nonlinear refractive index. This time-dependent effect becomes even more pronounced for longer pulses (>50 fs) where the system remains resonant over long time. It may therefore seem natural to investigate which transitions (e.g., bound-bound, bound-continuum, and continuum-continuum transitions as defined in \cite{Nurhuda1}) mostly contribute to the Kerr saturation and inversion in order to exhibit a simple physical interpretation about the sign inversion of the Kerr effect at high intensity. However, like the amount of ionized electrons in the presence of the external field, the bound-bound, bound-continuum, and continuum-continuum contributions to the total polarizability (i.e. the partial polarizabilities) are also gauge-dependent, as illustrated in Figs. \ref{Fig:3}(c,d). This prevents any rigorous identification of specific transitions responsible for the Kerr saturation and inversion \cite{Nurhuda1,Nurhuda2,Kano,Volkova}. More particularly, while only continuum-continuum transitions (i.e. the acceleration of electron in the continuum) seem to be responsible for the Kerr sign inversion in the length gauge, both continuum-continuum and bound-bound transitions actively participate in the velocity gauge. The fact that the partial nonlinear polarizability  $\Delta\alpha_{\textrm{CC}}$ induced by continuum-continuum transitions is a major negative contribution to the total nonlinear polarizability in both velocity and length gauges does not imply any similar behavior in other gauges. As a consequence, while the higher order Kerr effect controversy has crystalized the debate on whether or not ionized electrons (or equivalently the plasma) are the main source of the Kerr saturation and inversion mechanism, it appears that this question is in fact irrelevant due to the gauge-dependency of the answer. 

As a conclusion, time dependent non perturbative calculations show that the nonlinear polarizability of hydrogen saturates and reverses its sign around 30 TW/cm$^2$, an intensity at which refractive index inversion is observed \cite{LoriotHFL09,LaserPhysLoriot}. In this strong-field regime, characterized by ionization channel closure, the electrons promoted into the continuum significantly interact with their parent ions and cannot be treated as free charged particles during the pulse. As a consequence, the well admitted scenario of Kerr effect saturation and inversion misses a quasi-instantaneous supplementary negative term, which can be identified as higher order Kerr effect or alternatively as a deviation from the Drude model. This result is compatible with the original observations of Kerr saturation and inversion \cite{LoriotHFL09,LaserPhysLoriot}, even if plasma grating effects could have a partial contribution in the experimental observations. It is worth mentioning that a direct comparison with the birefringence experiments \cite{LoriotHFL09,LaserPhysLoriot} would call for an evaluation of the refractive index along two orthogonal directions. Such a calculation would need to include an electric field polarized along these two directions as well as magnetic sub levels transitions, which is beyond the scope of this work.  

Moreover, seeking to pinpoint specific transitions between atomic states (either bound or in the continuum) as responsible for the nonlinear refractive index inversion is irrelevant, since the intra-pulse populations (in particular, the free electron density or plasma density) are not physical observables, but rather gauge-dependent quantities. Further studies will be needed to exhibit a simple and gauge independent physical interpretation of the sign inversion of the Kerr effect. These results therefore draw a radically new perpective on the current debate about higher order Kerr effect. 

Although derived in atomic hydrogen, these conclusions should apply to all major air constituents since ionization channel closures occur in the same intensity regime in all of these gases. Further works will be needed to describe the Kerr effect with a ready-to-use, computationnally efficient parametrization. In particular, preliminary studies suggest that the full quantum trajectory of the electron could be described as those of a particle with a pulse intensity- and shape-dependent effective mass. This parametrization will be useful for practical purposes requiring repeated computation of the nonlinear refractive index, like the simulation of laser filamentation. 

Acknowledgments. This work was supported by the Conseil R\'egional de Bourgogne (FABER program), the \textit{FASTQUAST} ITN Program of the 7th FP. JPW acknowledges financial support from the ERC advanced grant "FilAtmo". EC and PB thank O. Peyrusse and the CRI-CCUB for CPU loan on their respective multiprocessor servers. 

\bibliographystyle{unsrt}

\end{document}